\documentclass[12pt,aps]{revtex4} 

\usepackage{graphicx}
\usepackage{amsmath} 
\usepackage{textcomp}
\usepackage{mathrsfs}
\usepackage{amsfonts}
\usepackage[draft]{hyperref}

\begin{document}

\newcommand{\vett}[1]{\mathbf{#1}}
\newcommand{\uvett}[1]{\hat{\vett{#1}}}
\newcommand{\beq}{\begin{equation}}
\newcommand{\eeq}{\end{equation}}
\newcommand{\barr}{\begin{eqnarray}}
\newcommand{\earr}{\end{eqnarray}}
\newcommand{\GH}{Goos-H\"anchen }
\newcommand{\IF}{Imbert-Fedorov }
\newcommand{\bra}[1]{\langle #1|}
\newcommand{\ket}[1]{| #1\rangle}
\newcommand{\expectation}[3]{\langle #1|#2|#3\rangle}
\newcommand{\braket}[2]{\langle #1|#2\rangle}

\title{\GH and \IF shifts for astigmatic Gaussian beams}

\author{Marco Ornigotti}
\affiliation{Institute of Applied Physics, Friedrich-Schiller University, Jena, Max-Wien Platz 1, 07743 Jena, Germany}
\email{marco.ornigotti@uni-jena.de}

\author{Andrea Aiello}
\affiliation{
Max Planck Institute for the Science of Light, G\"unther-Scharowsky-Strasse 1/Bau24, 91058 Erlangen, Germany}
\affiliation{Institute for Optics, Information and Photonics, University of Erlangen-Nuernberg, Staudtstrasse 7/B2, 91058 Erlangen, Germany }

\begin{abstract}
In this work we investigate the role of the beam astigmatism in the \GH and \IF shift. As a case study, we consider a Gaussian beam focused by an astigmatic lens and we calculate explicitly the corrections to the standard formulas for beam shifts due to the astigmatism induced by the lens. Our results show that astigmatism may enhance the angular part of the shift. \\
\end{abstract}
\maketitle
\section{Introduction}

Geometrical optics considers light fields as rays directed along the propagation direction of the field itself. Within this approximation, most of the phenomena that we witness daily regarding light can be easily explained. An example is given by the phenomenon of reflection from an interface, which for an optical ray happens in a specular way, following Snell's law \cite{ref1}. When the wave properties of optical fields are taken into account, however, deviations from specular reflection can be observed. This is the case of optical beams, whose finite transverse sizes affect its reflection and refraction across interfaces. The most common manifestation of these effects is given by the so-called \GH \cite{ref2,ref3,ref4} and \IF \cite{ref5,ref6,refIF1,refIF2,refIF3,refIF4,refIF5,refIF6,refIF7,refIF8,refIF9} shifts, the former occurring in the plane of incidence and the latter in the plane perpendicular to the plane of incidence. These phenomena have been extensively studied in the past for a vast category of beam configurations \cite{ref7,ref8,ref9,ref9a,ref13} and interfaces \cite{ref10a, ref10,ref11,ref12}. A comprehensive review on beam shift phenomena can be found in Ref. \cite{ref12a}. Recently, the analogy between beam shifts and the quantum mechanical weak measurements has been also pointed out \cite{ref13a,ref13b,res10,res11,res12}, which resulted in the possibility of observing amplified beam shifts \cite{merano1,merano2}. 

In this paper we theoretically investigate the effect of the beam astigmatism on both \GH and \IF shifts. Our calculations show that the astigmatism affects, at the leading order, both the spatial and angular shifts, with its main action being the introduction of an enhancement factor in the angular part of the shifts. The results presented here address the simple case of an astigmatic Gaussian beam, that we take as paradigmatic example for studying the effect of astigmatism on the \GH and \IF shifts.

\section{Model and Methods}
\begin{figure}[!h]
\begin{center}
\includegraphics[width=0.8\textwidth]{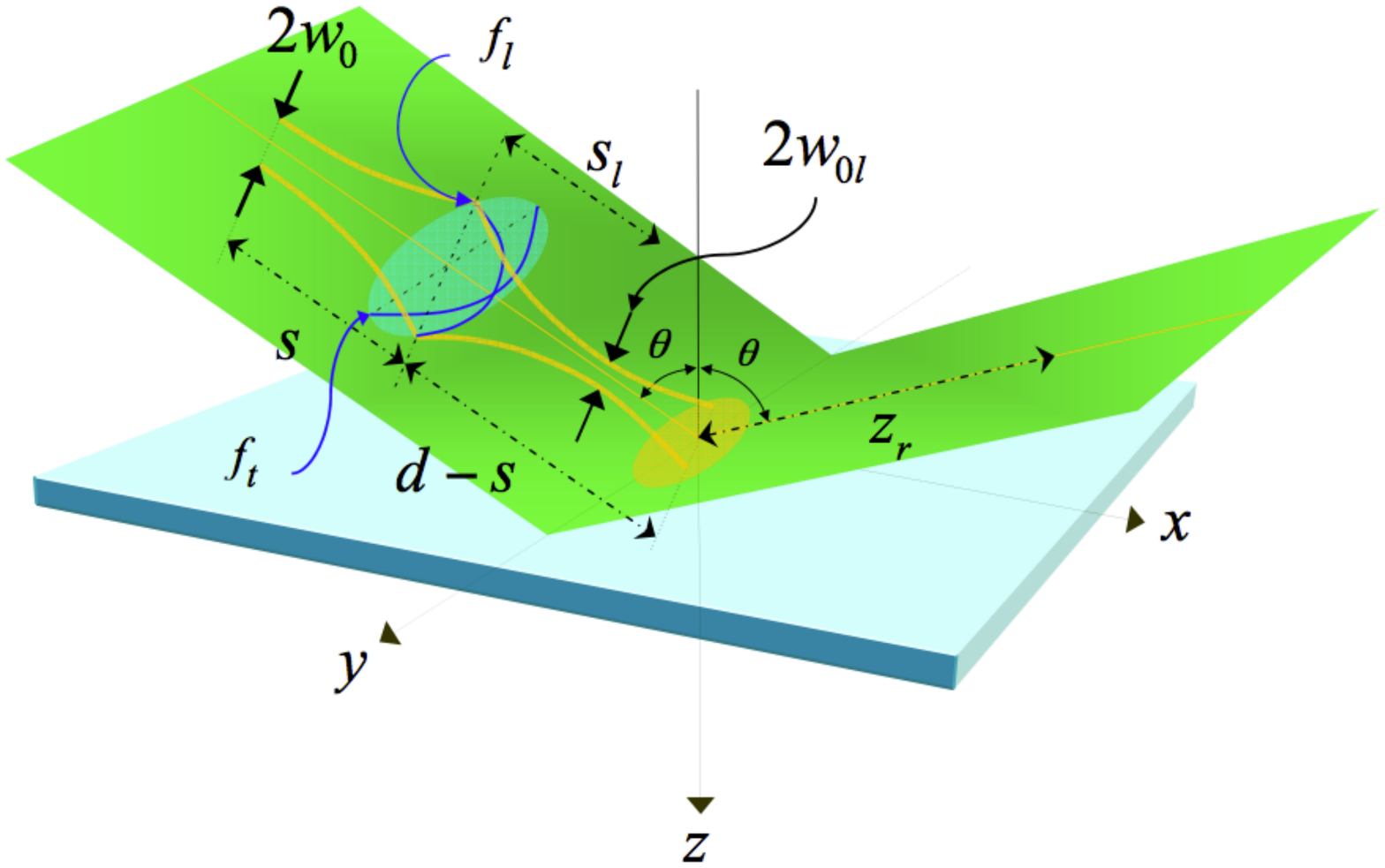}
\caption{Scheme of the beam reflection at the plane interface. Here $f_l$ and $f_t$ are the \emph{longitudinal} and \emph{transverse} focal lengths of the astigmatic lens, respectively. Note that for a cylindrical lens one has either $f_t=\infty$ or $f_l=\infty$. In addition, $s$ is the distance from the waist of the incident beam to the lens, and $d-s$ is the distance from the lens to the reflecting surface. $s_l$ is the distance from the lens to the longitudinal waist $w_{0l}$.  Similarly, we can define as $s_t$ the distance from the lens to the transverse waist $w_{0t}$ (not shown in the figure). Finally, $z_r$ is the distance from the reflecting surface to the detector position.}
\label{fig1}
\end{center}
\end{figure}

We start our analysis by considering a monochromatic fundamental Gaussian beam of frequency $\omega=c k$ (being $k=2\pi/\lambda$ the wave number) characterized by a beam waist $w_0$,  which impinges upon an astigmatic lens with two different focal lengths, namely $f_l$ in the longitudinal ($x$-) direction and $f_t$ in the transverse ($y$-) direction, as sketched in Fig. \ref{fig1}. After the lens, the beam will be focused at different lengths in the $x$ and $y$ direction. To calculate the beam waist in these directions, we make use of the well-known formula for paraxial beams \cite{svelto}:
\beq\label{eq0}
\frac{1}{q_{\mu}(s)}=\frac{1}{q(s)}-\frac{1}{f_{\mu}},
\eeq
where $\mu=\{l,t\}$, $f_{\mu}$ is the focal length, $s$ is the distance between the Gaussian beam's waist and the lens, and $q(s)=s-iz_R$ is the complex beam parameter, being $z_R=kw_0^2/2$ the Rayleigh range of the beam \cite{svelto} and $q_{\mu}(s+z)=q_{\mu}(s)+z$. Obtaining $q_{\mu}(s)$ from the previous equation and setting its real part to zero gives the distance $s_{\mu}$ from the lens to the new waist as
\beq
\frac{1}{s_{\mu}}=\frac{1}{f_{\mu}}-\frac{s-f_{\mu}}{s(s-f_{\mu})+z_R^2},
\eeq
while the imaginary part of $q_{\mu}$ allows us to write the longitudinal and transverse beam waist as follows:
\beq\label{astW}
\left(\frac{w_{0\mu}}{w_0}\right)^2=\frac{f_{\mu}^2}{(s-f_{\mu})^2+z_R^2}.
\eeq
According to the convention used in Refs. \cite{ref12a,ref14}, we describe the reflection process by means of three different reference frames, namely one attached to the incident beam ($\{x_i,y_i,z_i\}$), one attached to the reflected beam ($\{x_r,y_r,z_r\}$) and a third reference frame $\{x,y,z\}$, called the laboratory frame, attached to the reflecting interface. These reference frames are linked together by a rotation of an angle $\theta$ around the $y$ axis  \cite{praDuality}. The complex electric field impinging on the dielectric interface can be written in terms of its Fourier components, in the incident reference frame,  as follows:
\beq\label{first}
\vett{E}^I(\vett{r})=\sum_{\lambda=1}^2\int d^2K \vett{A}_{\lambda}(U,V,\theta)e^{ik[Ux_i+Vy_i+W(z_i+D)]},
\eeq
where $d^2K=dUdV$, $\theta$ is the angle of incidence and $U$,$V$ and $W$ are the components of the wave vector in the incident reference frame, defined by the relation
\beq
 \vett{k}=k(U\uvett{x}_i+V\uvett{y}_i+W\uvett{z}_i)=k_x\uvett{x}+k_y\uvett{y}+k_z\uvett{z}.
 \eeq
 
In Eq. \eqref{first}, $\vett{A}_{\lambda}(U,V,\theta)=\alpha_{\lambda}(U,V,\theta)A(U,V)\uvett{e}_{\lambda}(U,V,\theta)$ is the vector spectral amplitude of the field, with $\uvett{e}_{\lambda}(U,V,\theta)$ being the local reference frame attached to the incident field \cite{ref17}, $\alpha_{\lambda}(U,V,\theta)=\uvett{f}\cdot\uvett{e}_{\lambda}(U,V,\theta)$ are the projections of the initial beam polarization $\uvett{f}=f_p\uvett{x}+f_s\uvett{y}$ (with $|f_p|^2+|f_s|^2=1$) onto the local basis $\uvett{e}_1(\vett{k})$, $\uvett{e}_2(\vett{k})$ and $\uvett{k}$ attached to the wave vector $\vett{k}$, and $A(U,V)$ is the scalar spectral amplitude associated with the beam, that in our case assumes the form of an astigmatic Gaussian beam, i.e.,
\beq\label{astG}
A(U,V)=e^{-k^2(w_{0l}^2U^2+w_{0t}^2V^2)},
\eeq 
where $w_{0l}$ and $w_{0t}$ are defined according to Eq. \eqref{astW}. 

If we assume the beam to be well collimated, the paraxial approximation holds, and the expressions for the local reference frames and the polarization functions can be written in the following simple form \cite{ref17}:
\begin{subequations}
\begin{align}
\uvett{e}_1(U,V,\theta)&=\uvett{x}_i+V\cot\theta\uvett{y}_i-U\uvett{z}_i,\\
\uvett{e}_2(U,V,\theta)&=-V\cot\theta\uvett{x}_i+\uvett{y}_i-V\uvett{z}_i,
\end{align}
\end{subequations}
for the local basis, and
\begin{subequations}
\begin{align}
\alpha_1(U,V,\theta)&=f_P+f_SV\cot\theta,\\
\alpha_2(U,V,\theta)&=f_S-f_PV\cot\theta,
\end{align}
\end{subequations}
for the polarization functions. Notice, moreover, that the expression for $A(U,V)$ as given by Eq. \eqref{astG} has been written in correspondence of the beam waist after the lens. The beam, therefore, has propagated a distance 
$D=d-s-s_L$ (see Fig. \ref{fig1}) before reaching the reflection surface. This propagation factor is correctly taken into account by the $z$-dependent part of the angular spectrum of the incident field.

Upon reflection, the single plane wave components of the field described by Eq. \eqref{first} experience geometrical reflection, according to the Snell's law \cite{ref1}. The reflected electric field can be then represented by Eq. \eqref{first} with the substitution 
\beq
\uvett{e}_{\lambda}(\vett{k})e^{i\vett{k}\cdot\vett{r}}\rightarrow r_{\lambda}(\vett{k})\uvett{e}_{\lambda}(\tilde{\vett{k}})e^{i\tilde{\vett{k}}\cdot\vett{r}},
\eeq
where $r_{\lambda}(\vett{k})$ are the usual Fresnel coefficients for $p$ ($\lambda=1$) and $s$ ($\lambda=2$) polarization \cite{ref1}, and $\tilde{\vett{k}}=\vett{k}-2\uvett{z}(\uvett{z}\cdot\vett{k})=-U\uvett{x}_r+V\uvett{y}_r+W\uvett{z}_r$. The electric field in the reflected frame can now be defined as follows:
\beq\label{reflected}
\vett{E}^R(\vett{r})=\sum_{\lambda=1}^2\int d^2K \widetilde{\vett{A}}_{\lambda}(U,V)e^{i(-UX+VY+WZ')},
\eeq
where $X=kx_r$, $Y=ky_r$, $Z'=k(z_r+D)$ and the vector angular spectrum in the reflected frame is given by $\widetilde{\vett{A}}_{\lambda}(U,V)=A(U,V)r_{\lambda}(U,V)\alpha_{\lambda}(-U,V,\pi-\theta)\uvett{e}_{\lambda}(-U,V,\pi-\theta)$.  

The beam centroid can be then calculated as the weighted average of the position vector $\vett{R}=X\uvett{x}_r+Y\uvett{y}_r$ with respect to the total beam intensity in the reflected reference frame $|\vett{E}^R(x,y,z)|^2$. By employing the quantum notation for optical beams developed in Ref. \cite{quantumNot}, its explicit expression reads
\beq\label{centroid}
\langle\vett{R}\rangle=\frac{\expectation{\vett{E}^R}{\vett{R}}{\vett{E}^R}}{\braket{\vett{E}^R}{\vett{E}^R}}\equiv \langle X\rangle\uvett{x}_r+\langle Y\rangle\uvett{y}_r
\eeq
where $\langle \xi\rangle\equiv\langle \xi\rangle(z)$, with $\xi=X,Y$. The spatial ($\Delta$) and the angular ($\Theta$) \GH and \IF shifts are then given by the following relations:
\barr
\Delta_{GH}&=&\langle X\rangle|_{z=0},\hspace{1cm} \Theta_{GH}=\frac{\partial\langle X\rangle}{\partial z},\\
\Delta_{IF}&=&\langle Y\rangle|_{z=0}, \hspace{1cm} \Theta_{IF}=\frac{\partial\langle Y\rangle}{\partial z}.
\earr
By substituting into Eq. \eqref{centroid} the expression of the electric field in the reflected frame given by Eq. \eqref{reflected} with a first order accuracy in $(U,V)$ for the numerator and a second order accuracy for the denominator, we arrive, after a lengthy but simple calculation, to the following results:
\begin{subequations}\label{GHa}
\begin{align}
\Delta_{GH}&=\frac{1}{k}\left[\left(W_p\phi_p+W_s\phi_s\right)+\Gamma_x\left(W_p\rho_p+W_s\rho_s\right)\right],\\
\Theta_{GH}&=-\frac{1}{kz_R}\left(\frac{w_0}{w_{0l}}\right)^2\left(W_p\rho_p+W_s\rho_s\right),
\end{align}
\end{subequations}
for the \GH shifts, and
\begin{subequations}\label{IFa}
\begin{align}
\Delta_{IF}&=\frac{\sqrt{W_pW_s}\cot\theta}{k}\Big[\sin\xi+\frac{\Gamma_y(R_p^2-R_s^2)}{R_pR_s}\Big],\\
\Theta_{IF}&=-\frac{\sqrt{W_pW_s}}{kz_R}\left(\frac{w_0}{w_{0t}}\right)^2\frac{\left(R_p^2-R_s^2\right)}{R_pR_s}\cot\theta,
\end{align}
\end{subequations}
for the \IF shifts. In the previous equations, $\rho_{\lambda}=\operatorname{Re}\{\partial\ln r_{\lambda}/\partial\theta\}$, $\phi_{\lambda}=\operatorname{Im}\{\partial\ln r_{\lambda}/\partial\theta\}$, $R_{\lambda}=|r_{\lambda}|$ (with $\lambda\in\{p,s\}$) and 
\beq
\Gamma_{\mu}\equiv-\frac{D}{kw_{0\mu}^2}=-\frac{2\operatorname{Re}\{q_{\mu}(d)\}}{\operatorname{Im}\{q_{\mu}(d)\}},
\eeq
where $q_{\mu}(d)$ can be calculated from Eq. \eqref{eq0}. Moreover, $W_{\lambda}=f_{\lambda}^2R_{\lambda}^2/(f_p^2R_p^2+f_s^2R_s^2+\delta)$ is the fractional power contained in each polarization corrected by the beam's astigmatism factor $\delta$, whose explicit expression is given by
\barr\label{deltaA}
\delta&=&-\frac{1}{4k^2w_{0t}^2}\Big\{f_p^2\Big[\frac{w_{0t}^2}{2w_{0l}^2}\Big(\frac{\partial^2R_p^2}{\partial\theta^2}-2R_p^2\Big)+\cot^2\theta\Big(R_s^2-R_p^2\Big)\nonumber\\
&+&\cot\theta R_p\frac{\partial R_p}{\partial\theta}\Big]+f_s^2\Big[\frac{w_{0t}^2}{2w_{0l}^2}\frac{\partial^2R_s}{\partial\theta^2}+\cot^2\theta\Big(R_p^2-R_s^2\Big)\nonumber\\
&+&\cot\theta R_s\frac{\partial R_s}{\partial\theta}-R_s^2\Big]\Big\}.
\earr
Equations \eqref{GHa} and \eqref{IFa} are the first main result of this work. 

\section{Results and Discussion}

A closer inspection on Eqs. \eqref{GHa} and \eqref{IFa} reveals that the effect of the astigmatism on the \GH and \IF shifts is twofold. Firstly, the fractional power normalization factor $W_{\lambda}$ is changed by the quantity $\delta$. This means that, in contrast to the case of a fundamental Gaussian beam \cite{ref14}, the astigmatism introduces a correction on the energy stored in each polarization, being this correction dependent on both polarizations, as it appears clear from Eq. \eqref{deltaA}.  However, since this correction is of the second order into $(U,V)$, it is very small and can be therefore neglected. By doing so, therefore, the spatial shifts corresponds exactly to the ones of a fundamental Gaussian beam described in Ref. \cite{ref14}.

A second, and more interesting effect of the astigmatism can be observed in the angular shifts, where an extra multiplicative factor appears. By neglecting $\delta$ in the expression of $W_{\lambda}$, the angular shifts can be therefore rewritten in the following inspiring form:
\begin{subequations}
\begin{align}
\Theta_{GH}&=\left(\frac{w_0}{w_{0l}}\right)^2\Theta_{GH}^{(0)},\\
\Theta_{IF}&=\left(\frac{w_0}{w_{0t}}\right)^2\Theta_{IF}^{(0)},
\end{align}
\end{subequations}
\begin{figure}[!t]
\begin{center}
\includegraphics[width=\textwidth]{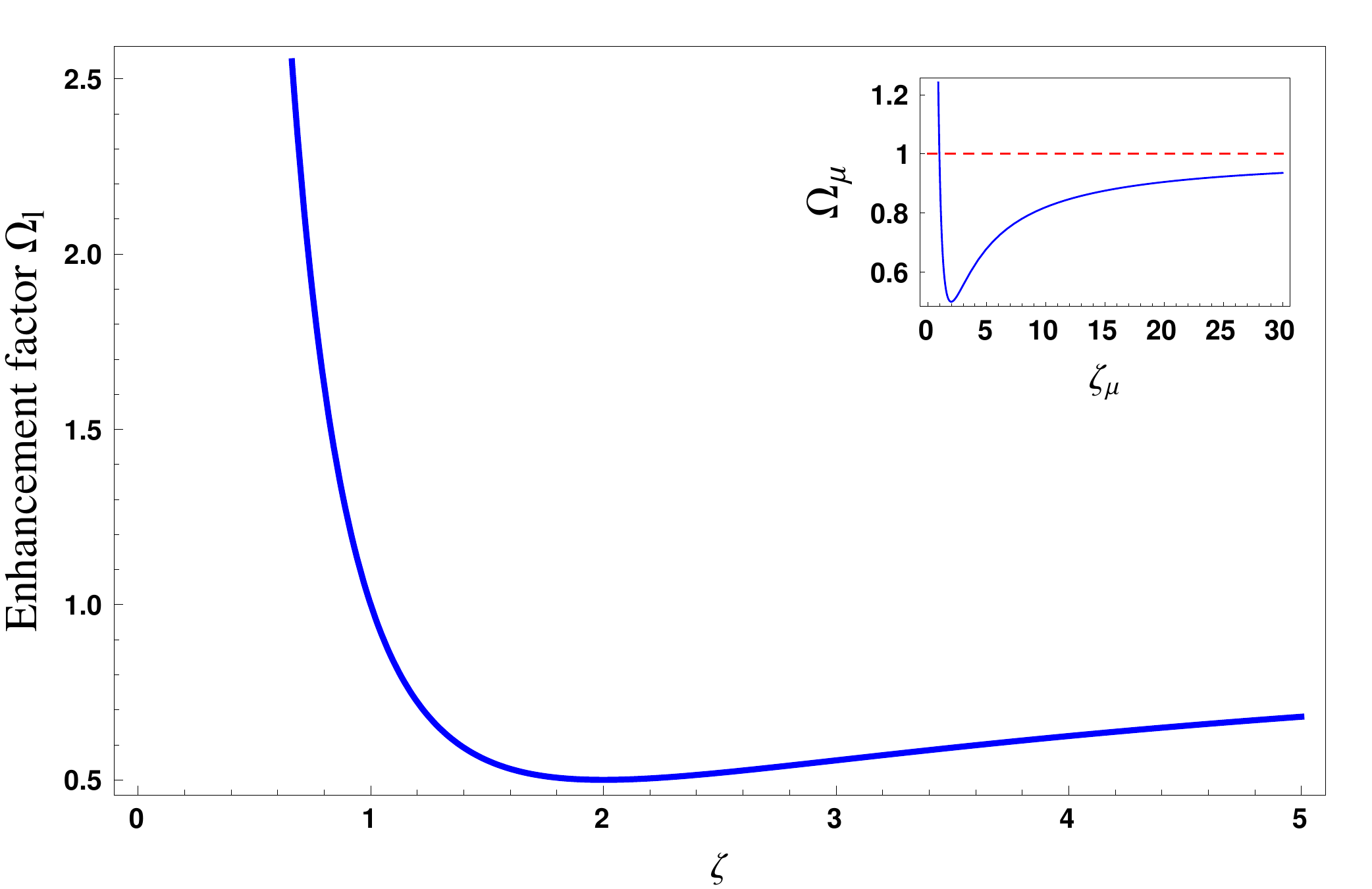}
\caption{Plot of the longitudinal enhancement factor $\Omega_l$ as a function of the normalized focal distance $\zeta=f_l/z_R$. As it can be seen, as $f_l\rightarrow 0$, the enhancement factor grows, thus providing a huge amplification of the angular shift. The same argument is valid for $\Omega_t$ as well, provided that the substitution $f_l\rightarrow f_t$ is understood.  The inset shows that when $\zeta_{\mu}\rightarrow\infty$ (i.e., when $f_{\mu}\rightarrow\infty$, with $\mu=\{l,t\}$), the enhancement factor $\Omega_{\mu}$ tends to unity. These plots have been obtained assuming $s/z_R=1$.}
\label{figure2}
\end{center}
\end{figure}
where  $\Theta_{GH,IF}^{(0)}$ refer to the angular \GH and \IF shifts for a fundamental Gaussian beam \cite{ref14}, respectively, and their explicit expression, in our case, is given by
\begin{subequations}
\begin{align}
\Theta_{GH}^{(0)}&=\frac{1}{kz_R}\left(W_p\rho_p+W_s\rho_s\right),\\
\Theta_{IF}^{(0)}&=-\frac{\sqrt{W_pW_s}}{kz_R}\frac{\left(R_p^2-R_s^2\right)}{R_pR_s}\cot\theta.
\end{align}
\end{subequations}
 This is our second main result. The astigmatism introduces an enhancement factor
\beq\label{omegaE}
\Omega_{\mu}=\left(\frac{w_0}{w_{0\mu}}\right)^2=\frac{(s-f_{\mu})^2+z_R^2}{f_{\mu}^2},
\eeq
 in the angular shifts that depends essentially on the degree of astigmatism of the beam in each direction, which in this case is here represented by the longitudinal or transversal focal length, respectively. 
 
 It is worth noticing, moreover, that the main effect of astigmatism is to introduce a \emph{selective} amplification of the angular GH and IF shifts, as it appears clear from Eqs. (18). According to those equations, in fact, while the amplification factor in front of the angular GH shift only contains the longitudinal focal length through the quantity $w_{0l}$, the transverse focal length only appears (through the term $w_{0t}$) in the angular IF shift.
 
 In order to quantify this enhancement, in Fig. \ref{figure2} $\Omega_l$ is plotted against the normalized focal length $\zeta=f_l/z_R$. As can be seen, for values of the focal length (either in the longitudinal or in the transversal direction, depending on which shift is one interested in) approaches zero, the enhancement factor diverges, thus giving a significative enhancement of the angular shift. Although a rigorous vectorial analysis of the focusing process for $f\rightarrow 0$ would reveal that the considered divergence is only an artifact of the paraxial equation, the essential feature of it, namely the amplification of the angular shifts  due to a shorter focusing length of the lens, is still very well captured by our paraxial model.
 
 \section{Conclusions}
In Conclusion, we have investigated the effect of a beam's astigmatism on the \GH and \IF shifts that occur when the beam impinges onto a dielectric surface. We have shown that the main effect of astigmatism is the introduction of an enhancement factor $\Omega_{\mu}$ on the angular shifts that is essentially depending on the degree of astigmatism of the beam in the longitudinal (transverse) direction. We have also shown that for small values of the focal length (compared with the beam's Rayleigh range $z_R$) of the lens we used to model the beam's astigmatism, a drastic amplification of the angular shifts occur. We believe that this result could be used experimentally to have another available channel of amplification, that could contribute to easily realize gigantic beam shifts that could lead to practical applications of this phenomenon.

 \section*{References}

\end{document}